\documentclass[english,11pt]{article}
\usepackage[english]{babel}
\usepackage{latexsym,amssymb,amsmath,amsfonts,epsfig,amsthm,color}
\usepackage{graphicx}
\usepackage{float}
\usepackage[left=1in,right=1in, top=1in, bottom=1in]{geometry}
\usepackage{framed}
\usepackage[leftbars,color]{changebar}
\usepackage{ifthen}
\usepackage{authblk}
\usepackage{colonequals}
\usepackage{array}
\usepackage[nottoc,numbib]{tocbibind}  
\usepackage{MnSymbol} 
\usepackage{braket}
\usepackage{cite}

\definecolor{webgreen}{rgb}{0,.5,0}
\definecolor{webblue}{rgb}{0,0,.5}
\usepackage[pdfpagelabels, colorlinks=true,urlcolor=webblue,linkcolor=webgreen,filecolor=webblue,citecolor=webgreen,pdfpagemode=UseOutlines,pdfstartview=FitH,pdfpagelayout=OneColumn,bookmarks=true,hyperfootnotes=false]{hyperref}
\usepackage[capitalise]{cleveref}
\makeatletter
\newtheorem*{rep@theorem}{\rep@title}
\newcommand{\newreptheorem}[2]{%
\newenvironment{rep#1}[1]{%
 \def\rep@title{#2 \ref*{##1}}%
 \begin{rep@theorem}}%
 {\end{rep@theorem}}}
\makeatother

\definecolor{darkgreen}{rgb}{0,.5,0}

\def\abs#1{\left| #1 \right|}

\newcommand{\eps}{\varepsilon}

\def\BibTeX{{\rm B\kern-.05em{\sc i\kern-.025em b}\kern-.08em
    T\kern-.1667em\lower.7ex\hbox{E}\kern-.125emX}}

\title{Quantum Cryptography \\
Beyond
Quantum Key Distribution}

\author{Anne Broadbent\footnote{Department of Mathematics and
    Statistics, University of Ottawa, Canada;
    \texttt{abroadbe@uottawa.ca}.}\,\, and Christian
  Schaffner\footnote{Institute for Logic, Language and Computation
    (ILLC), University of Amsterdam, and Centrum Wiskunde \&
    Informatica (CWI), The Netherlands; \texttt{c.schaffner@uva.nl}.}}

\date{}

\begin{document}

\maketitle

\begin{abstract}
Quantum cryptography is the art and science of exploiting quantum
mechanical effects in order to perform cryptographic
tasks. 
While the most well-known example of this discipline is quantum key
distribution (QKD), there exist many other applications such as
quantum money, randomness generation, secure two- and multi-party
computation and delegated quantum computation. Quantum
cryptography also studies the limitations and
challenges resulting from quantum adversaries---including the
impossibility of quantum bit commitment, the difficulty of quantum
rewinding and the definition of quantum security models for classical primitives.

In this review article, aimed primarily at cryptographers unfamiliar with the quantum world, we survey the area of theoretical quantum cryptography, with an emphasis on the constructions and limitations beyond the realm of QKD.

\bigskip

\noindent \textbf{Keywords:} Quantum cryptography $\cdot$  conjugate coding $\cdot$ quantum money $\cdot$ quantum key distribution $\cdot$ limited-quantum-storage models~$\cdot$ delegated quantum computation~$\cdot$ device-independence~$\cdot$ quantum bit commitment~$\cdot$ quantum two-party computations~$\cdot$ quantum rewinding~$\cdot$ superposition queries~$\cdot$ quantum random oracle model.

\bigskip

\noindent \textbf{Mathematics Subject Classification:} 94A60 $\cdot$ 
81P94 


\end{abstract}

\newpage
\tableofcontents

\newpage

\section{Introduction}

The relationship between quantum information and cryptography is almost half-a-century old: a 1968 manuscript of Wiesner (published more than a decade later~\cite{Wie83}), proposed \emph{quantum money} as the first ever application of quantum physics to cryptography,  and is also credited for the invention of oblivious transfer---a key concept in modern cryptography that was re-discovered years later by Rabin~\cite{Rab81}. Still today, the two areas are closely intertwined: for instance, two of the most well-known results in quantum information stand out as being related to cryptography: quantum key distribution (QKD)~\cite{BB84} and Shor's factoring algorithm~\cite{Sho94}.

There is no doubt that QKD has taken the spotlight in terms of the use of quantum information for cryptography  (in fact, so much that the term ``quantum cryptography'' is often \emph{equated} with QKD---a misconception that we aim to rectify here!); yet there exist many other uses of quantum information in cryptography.
 What is more, quantum information opens up the cryptographic landscape to allow functionalities that do not exist using classical\footnote{We use the word ``classical'' here and throughout to mean ``non-quantum''.} information alone, for example uncloneable quantum money.  We note, however, that the use of quantum information in cryptography has its limitations and challenges. For instance, we know that quantum information alone is insufficient to implement information-theoretically secure bit commitment; and that a   proof technique called \emph{rewinding} (which is commonly used in establishing a zero-knowledge property for a protocol) does not directly carry over to the quantum world and must re-visited in light of quantum information. \looseness=-1

In this paper, prepared on the occasion of the 25th anniversary edition of \emph{Designs, Codes and Cryptography}, we offer a survey of 
some of the most remarkable theoretical uses of quantum information for cryptography, as well as a number of limitations and challenges that cryptographers face in light of quantum information.  We assume that the reader is familiar with cryptography, but we do not assume any prior knowledge of quantum information. Quantum cryptography is a flourishing area of research, and we have chosen to give an overview of only a limited number of topics.
The reader is, of course, encouraged to follow up by consulting the references. To the best of our knowledge, prior survey  work on the topic of ``quantum cryptography beyond key exchange'' is limited to a   2006 survey by M{\"u}ller-Quade~\cite{Mul06} and a 1996 survey by Brassard and Cr\'epeau~\cite{BC96};
 see also an interesting personal account by Brassard~\cite{Bra05}. A number of surveys that focus on QKD exist and are listed in \cref{sec:QKD}.

\subsection{Overview}
\label{sec:intro-summary}

The predictions of quantum mechanics defy our everyday intuition: concepts such as \emph{superposition} (a particle can be in multiple places or states at the same time), \emph{entanglement} (particles are correlated beyond what is possible classically) and \emph{quantum uncertainty} (observing one property of a particle intrinsically degrades the possibility of observing another) are partly responsible for the bewildering possibilities in the quantum world.  \textbf{\cref{sec:basics}} of this survey contains a brief introduction to the mathematical formalism of quantum information as it pertains to quantum cryptography (no prior knowledge of quantum mechanics is assumed). Topics covered in this section include the mathematical formalism for the representation and manipulation of \emph{qubits} (the fundamental unit of quantum information). We also include a brief survey of concepts such as the quantum no-cloning theorem, entanglement and nonlocality --- all of which play an important role in quantum cryptography.

\textbf{\cref{sec:constructions}} of this survey is devoted to quantum cryptographic constructions.
The principal appeal in using 
quantum information for cryptography is in establishing a \emph{qualitative} advantage.
More precisely, the goal is to develop quantum cryptographic protocols that achieve some functionality in a way that is fundamentally advantageous compared to using classical information alone. This quantum advantage can be of the following types:
\begin{itemize}
 \item  a quantum protocol achieves information-theoretic (statistical) security; any classical protocol achieves this task with computational security at best;
 \item  a quantum protocol achieves computational security; no classical protocol can achieve this task, even with computational security.
\end{itemize}

 Many of the quantum constructions that we cover in this survey (whether they are of the first or second type) are inspired by the original proposal of Wiesner called \emph{conjugate coding}. This construction embodies many unique features of quantum information, as we explain in \cref{sec:conj-coding}. In particular, conjugate coding  is used in constructions for physically unforgeable quantum money (\cref{sec:q-Money}), as well as in quantum key distribution---a method that allows the information-theoretically secure expansion of shared keys (\cref{sec:QKD}). Another application of conjugate coding in quantum cryptography is in showing that two basic cryptographic primitives, \emph{bit commitment} and \emph{oblivious transfer} are (information-theoretically) \emph{equivalent} in the quantum world (\cref{sec:BC=>OT})---an equivalence that is provably false in the classical world.  Technologically speaking, perfect quantum communication and storage is a challenge; in building protocols in the \emph{bounded- and limited-storage quantum models} (collectively known as \emph{limited-storage models}),  ingenious cryptographers have turned this challenge to their advantage (\cref{sec:Bounded-storage})---once again, the key ingredient in the construction being conjugate coding. Motivated by the perspective that quantum computations will, in the future, be outsourced to remote locations (again, because of technological challenges involved in building quantum computers), cryptographers have studied protocols for the \emph{delegation of quantum computations}, which we cover in \cref{sec:delegatedQC}.  In \cref{sec:weak-coin-flip}, we review the possibility of quantum primitives that accomplish a security against malicious participants that is typically too weak for cryptographic applications (since it does not provide \emph{exponential} security), yet is still of interest due to the advantage that quantum information provides: this is embodied in quantum protocols for \emph{weak coin flipping} and \emph{imperfect bit commitment}. Finally, we survey \emph{device-independent cryptography} (\cref{sec:deviceindependence}) which can be seen as a culmination of many of the constructions already mentioned: thanks to this sophisticated technique, it is possible to achieve cryptographic tasks such as quantum key distribution and randomness expansion/amplification, with \emph{untrusted devices}, which are quantum devices that are assumed to have originated from an adversary. The very possibility of  achieving this result stems from one of the most mysterious quantum phenomena, namely \emph{nonlocality} (which is introduced in \cref{sec:basics-entanglement}).

While quantum information provides a number of advantages for cryptography, it also has its unique limitations and challenges, which we survey in \textbf{\cref{sec:limitations}}.
The first limitations that we survey are in terms of \emph{impossibility} results, namely the impossibility of information-theoretically secure quantum bit commitment (\cref{sec:impos-bit-commitment}) and of information-theoretically secure two-party quantum computation (\cref{sec:2PartyImpossible}).  Next, we cover two topics that  are applicable to purely classical protocols, in which essentially the only concern is that the adversary is capable of quantum information processing: \emph{quantum rewinding} (\cref{sec:quantum-rewinding}) and \emph{superposition attacks} (\cref{sec:quantum-random-oracles}). We emphasize that the cryptographic challenges encountered here are not related to the superior computational power of a quantum adversary, but rather stem from  quantum phenomena such as the no-cloning theorem (which forces us to develop an  alternative to the common \emph{rewinding} method used in order to establish the zero-knowledge property of interactive protocols), and of quantum superposition (which requires a new framework describing interactions with oracles---namely in the \emph{quantum random oracle model}).  Finally, in \cref{sec:position-based}, we survey the research area of position-based quantum cryptography, where players use their geographical position as cryptographic credential; while the current main result in this area is a no-go theorem for quantum protocols for the task of position verification, the \emph{possibility} of position-based quantum cryptography against \emph{resource-bounded} adversaries remains a tantalizing open question.

One of the lessons learned from all these impossibility results is that quantum security is definitely a tricky business---quantum cryptographers should take this as a warning:
the desire to find a ``quantum advantage'' (\emph{i.e.}~an application where quantum information outperforms all classical solutions) is extremely strong, and cryptographers must be vigilant since the quantum world comes with an abundance of subtleties.

\medskip
We note that the bibliographic entries in this survey are available as an open-source \BibTeX\ file at \url{https://github.com/cschaffner/quantum-bib}.

\subsection{Further topics}
\label{sec:further-topics}
Already, the literature on quantum cryptography is vast, and in this survey we have chosen to focus on only a few topics. We briefly mention here some topics that are \emph{not} included in this survey:

\begin{itemize}
\item \textbf{Everlasting security.} A protocol has everlasting security if it is secure against adversaries that are computationally unlimited \emph{after} the protocol execution. This type of security is very difficult to obtain classically, even under strong setup assumptions such as a common reference string or signature cards~\cite{Unr13}. Even though we do not treat the notion explicitly in this survey, many quantum-cryptographic protocols such as QKD (\cref{sec:QKD}) and the limited-quantum-storage protocols (\cref{sec:Bounded-storage}) come with the important benefit of everlasting security. In fact, everlasting security might be the most important reason to use QKD in the first place~\cite{SML10}. 
 \item \textbf{Quantum functionalities.} In this survey, we focus mainly on classical functionalities, but \emph{quantum} functionalities are of course also of interest: assuming full quantum computers for all parties, one can study the secure realization of a quantum ideal functionality. This includes topics such as the encryption of quantum messages in the information-theoretic \cite{AMTW00,BR03,HLSW04,Leu02}, entropic \cite{Des09,DD10} and computational~\cite{BJ15} settings, quantum secret sharing~\cite{CGL99},
 multi-party quantum computation \cite{BCG+06},  authentication of quantum messages~\cite{BCG+02},  two-party secure function evaluation~\cite{DNS10,DNS12},  quantum anonymous transmission~\cite{CW05,BBF+07},  quantum one-time programs \cite{BGS13} and quantum homomorphic encryption~\cite{YPF14,BJ15}.
   \item \textbf{Key recycling.}  Using quantum information, it is possible to detect eavesdropping such that key re-use is possible (if no eavesdropping is detected), while maintaining information-theoretic security. This idea was originally proposed by Bennett, Brassard and Breidbart in 1982~\cite{BBB14} and was worked out in detail by Damg{\aa}rd, Pedersen and Salvail in~\cite{DPS14}.
\item \textbf{Quantum uncloneability.} Because quantum information cannot, in general, be duplicated, (see~\cref{sec:no-cloning}), we can achieve functionalities related to copy-protection that cannot be obtained in the classical world. These include  uncloneable encryption~\cite{Got03}, quantum copy-protection~\cite{Aar09} and  revocable time-release encryption~\cite{Unr14}.
\item \textbf{Isolation assumptions.}  The \emph{multi-prover} interactive proof scenario~\cite{BGKW88} enables the information-theoretic implementation of primitives that are unachievable in the single-prover setting. However, the study of quantum information has shed new light on the ``isolation'' assumptions that are required in order to establish security~\cite{CSST11}. Related to this is the study of protocols which are secure against provers sharing correlations that are very strong, yet do not allow signalling:~\cite{KRR14,FF15}.
One particular way to enforce  isolation is to spatially separate the players by a far enough distance: the relativistic no-signalling principle ensuring that no information can travel faster than the speed of light between the two sites. A relativistic (classical) bit-commitment scheme was first proposed by Kent~\cite{Ken99} and lately improved and experimentally implemented~\cite{LKB+15}. See also related work~\cite{LKB+13}.
\item \textbf{Leakage resilience using quantum techniques.}  In \emph{leakage resilient computation}, we are interested in protecting computations from attacks according to various leakage models. In one of these models (the ``split-state'' model), it was shown~\cite{DDN15} that quantum information allows a solution to the \emph{orthogonal-vector problem}, while no classical solution exists. Also, related work~\cite{LRR14arxiv} shows that techniques from fault-tolerant quantum computation can be used to construct novel leakage-resilient classical protocols.
\item \textbf{Quantum cryptanalysis.} Historically, the study of quantum algorithms is closely related to  \emph{quantum cryptanalysis}, which is the study of quantum algorithms for cryptanalysis.
This is evidenced by some of the very early work on quantum algorithms, including Shor's algorithm for computing discrete logarithms and integer factoring in quantum polynomial-time~\cite{Sho94}, Grover's search algorithm~\cite{Gro96,BBHT98} (which provides a square-root speedup in term of query complexity,  for searching in an unstructured database).
Recent work in the area of quantum cryptanalysis includes~\cite{Reg04,Hal05,Hal07,BJLM13,CJS13,LMvdP13}.  See also \cite{Mos09,BvD10} for surveys on quantum algorithms, as well as the \emph{Quantum Algorithm Zoo}.\footnote{\url{http://math.nist.gov/quantum/zoo/}}

\item\textbf{Merkle puzzles in a quantum world.} The first unclassified proposal for \emph{secure communication over insecure channels} was made by Merkle in 1974 (published years later~\cite{Mer78}). The main idea is that honest parties can establish a secure communication channel by expending work proportional to~$N$, yet any successful attack requires a computational effort proportional to~$N^2$. In a nutshell, Grover's search algorithm~\cite{Gro96} implies that quantum computers break the security of Merkle's scheme. However, \cite{BHK+11} show how to restore security in the quantum context: either by using a new classical protocol (in which case an adversary can break the scheme by expending work proportional to $N^{5/3}$ (which is shown to be optimal), or by using a quantum protocol (in which case the quadratic security can essentially be restored).

\item \textbf{From classical to quantum security.} What can we say about the relationship between security in the classical setting versus the quantum setting? In this context,  Unruh~\cite{Unr10} shows that if a protocol is statistically secure in the universal composability (UC) framework~\cite{Can01,Unr10}, then the same protocol is quantum UC secure as well, and Fehr, Katz, Song, Zhou and Zikas~\cite{FKS+13}  classified the feasibility of cryptographic functionalities in the universal composability (UC) framework\footnote{A primitive is feasible if it can be implemented in the UC model from secure channels only.}, and showed that feasibility in the quantum world is equivalent (for a large family of functionalities) to classical feasibility, both in the computational and statistical setting. See also \cite{HSS11,Son14}.

  \item \textbf{Post-quantum cryptography.} The area of \emph{post-quantum cryptography}~\cite{BBD09}\footnote{The term is quite well-established by now, but chosen somewhat unfortunately, because the research area is concerned with cryptography which is still secure \emph{at the beginning} and not after the end of the era of large-scale quantum computers.}  finds alternatives to the RSA and discrete-log assumptions in (classical) cryptography, in order to circumvent quantum attacks stemming from Shor's algorithm. This area is traditionally considered a topic related to classical cryptography, but the issues arising from the problem of quantum rewinding (\cref{sec:quantum-rewinding}) and the superposition model for oracle access (\cref{sec:quantum-random-oracles}) may be considered part of post-quantum cryptography as well.

\item \textbf{Quantum public keys.} The use of quantum states as public keys presents fundamental challenges in terms of verifiability and re-usability; nevertheless, these types of keys have been studied in the context of \emph{quantum digital signatures}~\cite{GC01,DWA14} and quantum public-key identification~\cite{IM14}.

  \item \textbf{Distributed quantum computation.} Quantum information in known to provide advantages in distributed computation, namely in terms of quantum Byzantine agreement~\cite{BH05} and a quantum leader election in anonymous networks~\cite{TKM12}.

\item \textbf{Experimental implementations.} Experimental implementations of quantum cryptography are mostly focused on QKD (see \cite{ABB+14}), but also include   quantum coin flipping~\cite{MVUZ05,NFHM08,BBB+11}, quantum secret sharing~\cite{TZG01}, delegated quantum computation~\cite{BKB+12,FBS+14},  limited-quantum-storage cryptography~\cite{NJM+12,ENG+14}, and
     device-independent randomness generation~\cite{PAM+10,CMA+13}.
\end{itemize}


\section{Basics of Quantum Information}
\label{sec:basics}

This section contains the rudiments of quantum information that are used in the main text; we assume of the reader only basic knowledge of linear algebra.
The reader should be warned that quantum theory is actually much more rich, subtle and beautiful! Textbook references on quantum information include: \cite{NC00,KLM07,Mer07,Wil13,Wat15}.

\subsection{State Space}
The \emph{bit} is the fundamental unit of information for classical information processing. In quantum information processing, the corresponding unit is the
\emph{qubit}, which is  described mathematically by a vector of length one in a two-dimensional complex vector space. We use notation from physics to denote vectors that represent quantum states, enclosing vectors in a \emph{ket}, yielding, \emph{i.e.}~$\ket{\psi}$. We can write any state on one qubit as a $\ket{\psi} = \alpha\ket{0} + \beta\ket{1}$, where the states $\ket{0}$ and $\ket{1}$ form a basis for the underlying two-dimensional vector space, and where $\alpha$, $\beta$ are complex numbers satisfying $|\alpha|^2 + |\beta|^2 =1$. If neither $\alpha$ nor $\beta$ are zero, then we say that~$\ket{\psi}$ is in a \emph{superposition} (linear combination) of both $\ket{0}$ and $\ket{1}$.
The quantum state of two or more qubits can be described by a tensor product. Hence, the four basis states for two qubits are $\ket{0} \otimes \ket{0}, \ket{0} \otimes \ket{1}, \ket{1} \otimes \ket{0}, \ket{1} \otimes \ket{1}$ which is usually abbreviated as $\ket{00},\ket{01},\ket{10},\ket{11}$.
Extending  the concept of superposition to multiple qubits, we see that a system of~$n$ qubits can be in any superposition of the $n$-bit basis states $\ket{00\ldots0}, \ket{00\ldots1}, \ldots, \ket{11\ldots1}$. Hence, an $n$-qubit state is described by $2^n$ complex coefficients.
In case of \emph{bipartite} quantum states shared among Alice and Bob, subscripts can be used to indicate which player holds which qubits. For instance, the 2-qubit state $\ket{0}_A \otimes \ket{0}_B = \ket{00}_{AB}$ means that Alice and Bob both hold a qubit in state $\ket{0}$.

\subsection{Unitary Evolution and Circuits}
Basic evolutions of a quantum system are described by linear operations that preserve the norm; formally, these operations can be expressed as \emph{unitary} complex matrices (a complex matrix~$U$ is \emph{unitary} if ${UU}^\dag =\mathbb{I}$, where ${U}^\dag$ is the complex-conjugate transpose of~$U$). Quantum algorithms are commonly described as circuits (rather than by quantum Turing machines) consisting of basic quantum gates from a universal set. Commonly used single-qubit gates are the negation (${X}$), phase (${Z}$) and Hadamard (${H}$) gates, expressed by the following unitary matrices:
\begin{equation}
{X}= \begin{pmatrix} 0 & 1 \\ 1 & 0 \end{pmatrix}, \,\, {Z}= \begin{pmatrix} 1 & 0 \\0 & -1 \end{pmatrix}, \,\, {H}= \frac{1}{\sqrt{2}}\begin{pmatrix} 1 & 1 \\1 & -1 \end{pmatrix}\,.
\end{equation}
An example of a two-qubit gate is the controlled-not operation (${CNOT}$):
\begin{equation}
{CNOT}= \begin{pmatrix} 1 & 0 & 0 &0  \\ 0 & 1 & 0 &0 \\0 & 0 & 0 &1  \\ 0 & 0 & 1 &0 \end{pmatrix}\,.
\end{equation}

\subsection{Measurement}
In addition to unitary evolution, we specify an operation  called \emph{measurement}, which, in the simplest case, takes a qubit and outputs a classical bit.
If we measure a qubit $\ket{\psi} = \alpha\ket{0} + \beta\ket{1}$, we will get as outcome a single bit, which takes the value~0 with probability~$\abs{\alpha}^2$ and the value~1 with probability~$\abs{\beta}^2$. We further specify that, after the process of measurement, the quantum system \emph{collapses} to the measured outcome. Thus, the quantum state is \emph{disturbed} and it becomes \emph{classical}: any further measurements have a deterministic outcome.  We have described measurement with respect to the \emph{standard basis}; of course, a measurement can be described according to an arbitrary basis;
the probabilities of the outcomes can be computed by first applying the corresponding change-of-basis, followed by the standard basis measurement.
Measurements can actually  be described much more generally: \emph{e.g.}\ we can describe outcomes of measurements of a strict subset of a quantum system---the mathematical formalism to describe the outcomes  uses the \emph{density matrix} formalism, which we do not describe~here.

As a simple example of quantum measurements, consider the states $\ket{\psi_1} = \ket{0}$ and $\ket{\psi_2} = \frac{1}{\sqrt{2}}(\ket{0} + \beta\ket{1})$. Then measuring the state $\ket{\psi_1}$  yields the outcome~$1$ with unit probability and the state remains $\ket{0}$, while measuring $\ket{\psi_2}$ yields the outcome 0 or 1, each with probability~$\frac{1}{2}$, and the post-measurement state is $\ket{0}$ if we observed outcome 0 and $\ket{1}$ if we observed $1$.

\subsection{Quantum No-Cloning}
\label{sec:no-cloning}
One of the most fundamental properties of quantum information is that it is not physically possible, in general, to \emph{clone} a 
quantum system~\cite{WZ82} (\emph{i.e.}\ there is no physical process that takes as input a single quantum system, and outputs two identical copies of its input). A simple proof follows from the linearity of quantum operations\footnote{Assume a quantum operation $A$ which takes as input a qubit in state $\ket{\psi}$ (together with a ``helping'' qubit in state $\ket{0}$) and outputs $\ket{\psi}\ket{\psi}$. Hence, $A \ket{0}\ket{0} = \ket{0}\ket{0}$ and $A \ket{1}\ket{0} = \ket{1}\ket{1}$. By linearity of $A$, it must hold that $A \frac{\ket{0}+\ket{1}}{\sqrt{2}} \ket{0} = \frac{A \ket{0}\ket{0} + A \ket{1}\ket{0}}{\sqrt{2}} = \frac{\ket{0}\ket{0} + \ket{1}\ket{1}}{\sqrt{2}}$ which is not equal to the state $\frac{\ket{0}+\ket{1}}{\sqrt{2}} \otimes \frac{\ket{0}+\ket{1}}{\sqrt{2}}$ which we would expect as output from a perfect copying operation $A$. Hence, such an $A$ does not exist.}.
 At the intuitive level, this principle is present in almost all of quantum cryptography, since it prevents the classical reconstruction of the description of a given qubit system. For instance, given a single copy of a general qubit $ \alpha\ket{0} + \beta\ket{1}$, it is not possible to ``extract'' a full classical description of $\alpha$ and $\beta$, because measuring disturbs the state. At the formal level, however, we generally require more sophisticated tools to prove the security of quantum cryptography protocols (see \cref{sec:conj-coding}).

\subsection{Quantum Entanglement and Nonlocality}
\label{sec:basics-entanglement}

A crucial and rather counter-intuitive feature of quantum mechanics is \emph{quantum entanglement}, a physical phenomenon that occurs when quantum particles behave in such a way that the quantum state of each particle cannot be described individually.
A simple example of such an entangled state are two qubits in the state $(\ket{00}_{AB}+\ket{11}_{AB})/\sqrt{2}$. When Alice measures her qubit (in system~$A$), she obtains a random bit $a \in \{0,1\}$ as outcome and her qubit collapses to the state $\ket{a}_A$ she observed. At the same time, Bob's qubit (in system~$B$) also collapses to $\ket{a}_B$ and hence, a subsequent measurement by Bob yields the same outcome $b=a$. It is important to realize that this collapse of state at Bob's side occurs simultaneously with Alice's measurement, but it does not allow the players to \emph{send information} from Alice to Bob. It simply provides Alice and Bob with a shared random bit. In general, quantum entanglement does not contradict the fundamental \emph{non-signaling principle} of the theory of relativity stating that no information can travel faster than the speed of~light.\footnote{However, this ``spooky action at a distance'' puzzled Einstein a lot, and was the inspiration for his very influential paper co-authored with Podolsky and Rosen~\cite{EPR35}. Nowadays, we often call two qubits in the maximally entangled state $(\ket{00}_{AB}+\ket{11}_{AB})/\sqrt{2}$ an \emph{EPR pair}.}

It turns out that by measuring entangled quantum states, Alice and Bob are able to produce correlations that are stronger than all correlations they could obtain when sharing only classical randomness. In this case, physicists say that the correlations \emph{violate a Bell inequality}~\cite{Bel64}. The most well-known example of such an inequality was proposed by Clauser, Horne, Shimony and Holt~\cite{CHSH69}. It can be described as a so-called \emph{non-local game} among two players Alice and Bob. In this CHSH game, Alice and Bob can initially discuss in order to establish a joint strategy. Once the game starts, they are separated and cannot communicate. They receive as input uniformly random bits $x$ and $y$ and have to output bits $a$ and $b$ respectively. They win the game if and only if $a \oplus b = x \wedge y$ (imagine a third party, called a \emph{referee} who chooses $x$ and $y$, receives $a$ and $b$ and checks whether the relationship $a \oplus b = x \wedge y$ holds). A possible classical strategy for Alice and Bob is to ignore their inputs and always output $a=b=0$. This strategy lets them win the game with probability $3/4$. It can be checked that there exist no better strategy for two classical players who are not allowed to communicate. In other words, we have the Bell inequality $\Pr[\mbox{classical players win CHSH}] \leq 3/4$. However, if Alice and Bob share a maximally entangled state (e.g.\ an EPR pair $(\ket{00}+\ket{11})/\sqrt{2}$), they can perform a quantum measurement which allows them to win the CHSH game with probability $\cos^2(\pi/8) \approx 0.85$ which is strictly larger than $3/4$, hence violating the Bell inequality. Many experimental tests of this inequality have been performed and consistently found violations of this inequality, thereby proving that the world is actually more accurately described by quantum mechanics rather than by classical mechanics.

\subsection{Physical Representations}
The mathematical model of quantum mechanics is currently the most accurate description of the physical world. This theory is without doubt the most successful and well-tested physical theory of all times; it describes a wide range of physical systems, and thus offers a large number of possible physical systems which can serve as quantum devices. These possibilities include
photonic quantum computing, superconduction qubits, nuclear magnetic resonance, ion trap quantum computing and atomic quantum computing (see, \emph{e.g.}~\cite{dMM03}). For our purposes, at the abstract level, all of these systems are described by the same formalism; however there may be experimental reasons to prefer one implementation over the other (\emph{e.g.}\ photons are well-suited for long-distance quantum communications, but other systems such as superconducting qubits are better suited for quantum interactions.)


\section{Quantum Cryptographic Constructions}
\label{sec:constructions}

In this section, we survey a number of quantum cryptographic protocols (see \cref{sec:intro-summary} for a brief overview of these topics). Many of these protocols share the remarkable feature of being based on a very simple pattern of quantum information called \emph{conjugate coding}. Because of its paramount importance in quantum cryptography, we first present this notion in \cref{sec:conj-coding}. We then show how conjugate coding is the crucial ingredient in the quantum-cryptographic constructions for quantum money (\cref{sec:q-Money}), quantum key distribution~(\cref{sec:QKD}), a quantum reduction from oblivious transfer to bit commitment (\cref{sec:BC=>OT}), the limited-quantum-storage model (\cref{sec:Bounded-storage}) and delegated quantum computation~(\cref{sec:delegatedQC}). Further topics covered in this section are quantum coin-flipping (\cref{sec:weak-coin-flip}) and device-independent cryptography (\cref{sec:deviceindependence}).

\subsection{Conjugate Coding}
\label{sec:conj-coding}
 \emph{Conjugate coding}~\cite{Wie83}
is based on the principle that we can encode \emph{classical information}  into \emph{conjugate} quantum bases. This primitive is extremely important in quantum cryptography---in fact, the vast majority  of quantum cryptographic protocols exploit conjugate coding in one way or another. Conjugate coding is also called \emph{quantum coding}~\cite{BBB14} and \emph{quantum multiplexing}~\cite{BBBW82}.

The principle of conjugate coding is simple: for clarity of presentation and consistency with commonly used terminology, we associate a qubit with a photon (a particle of light), and use photon polarization as a quantum degree of freedom. Among others, photons can be polarized horizontally $(\ket{\leftrightarrow})$, vertically ($\ket{\updownarrow}$), diagonally to the left ($\ket{\nwsearrow}$), or diagonally to the right  ($\ket{\neswarrow}$). Photon polarization is a quantum property, and by associating $ \ket{\leftrightarrow} = \ket{0}$,$\ket{\updownarrow} = \ket{1}$, $\ket{\nwsearrow}= \frac{1}{\sqrt{2}}(\ket{0} + \ket{1})$ and $\ket{\neswarrow}=\frac{1}{\sqrt{2}}(\ket{0} - \ket{1})$, we can apply  quantum operations to these states, as in \cref{sec:basics}.

Each set $R=\{\ket{\leftrightarrow},\ket{\updownarrow}\}$ and $D=\{\ket{\nwsearrow},\ket{\neswarrow}\}$ forms a basis (called the \emph{rectilinear} and \emph{diagonal} bases, respectively), and can thus be used to encode a classical bit (see \cref{tbl:conj-coding}). $R$ and $D$ are \emph{conjugate bases}.

\begin{table}[h]
\centering
\begin{tabular}{lcccccccccc}
\hline
encoded bit        & 0              & 0            & 1            & 0            & 1                 & 1                 & 0            & 0              & 0              & 0 \\
basis choice   & $R$              & $D$            & $D$            & $D$            & $R$                 & $R$                 & $D$            & $R$              & $R$              & $D$ \\ \hline
quantum encoding & $\ket{\updownarrow}$ & $\ket{\neswarrow}$ & $\ket{\nwsearrow}$ & $\ket{\neswarrow}$ & $\ket{\leftrightarrow}$ & $\ket{\leftrightarrow}$ & $\ket{\neswarrow}$ & $\ket{\updownarrow}$ & $\ket{\updownarrow}$ & $\ket{\neswarrow}$  \\ \hline
\end{tabular}
\caption{Example of conjugate coding. Here, we use the abbreviation $R$ for the \emph{rectilinear} basis and $D$ for the \emph{diagonal} basis.
\label{tbl:conj-coding}}
\end{table}

 The relevance of conjugate coding to cryptography is summarized by two key features that were, remarkably, already mentioned and exploited in Wiesner's  work~\cite{Wie83}:
 \begin{enumerate}
 \item  Measuring in one basis irrevocably destroys any information about the encoding in its conjugate basis.
 \item  The originator of the quantum encoding can verify its authenticity; however, without knowledge of the encoding basis, and given access to a \emph{single} encoded state, no third party can create two quantum states that pass this verification procedure with high probability.
 \end{enumerate}

In order to explain the first property, recall the well-known \emph{Heisenberg uncertainty relation}~\cite{Hei27}, which forbids learning both the position and momentum of a quantum particle precisely and simultaneously. In terms of photon polarization, and for a single photon, let us denote by $P_X$ the distribution of outcomes when measuring the photon in the rectilinear  basis and by~$Q_X$ the distribution when measuring in the diagonal basis. Following Heisenberg, Maassen and Uffink~\cite{MU88} showed an \emph{uncertainty relation}: $H(P_X) + H(Q_X) \geq 1$ (where $H$ is the \emph{Shannon entropy}, an information-theoretic measure of uncertainty given by  $H(P_X) = -\sum_x p_x \log_2p_x$).
 Intuitively, such a relation quantifies the fact that one can know the outcome exactly in one basis, but consequently has complete uncertainty in the other basis. Looking ahead, we will see that such 
 uncertainty relations play a key role in proving security of quantum cryptographic protocols, \emph{e.g.}~in the limited-quantum-storage setting (\cref{sec:Bounded-storage}). The second property above is explained by noting that a quantum encoding can be verified by measuring each qubit in its encoding basis and checking that the measurement result corresponds to the correct encoded bit. Intuitively, the \emph{no-cloning} theorem (\cref{sec:no-cloning}) prevents a third party from \emph{forging} a state that would pass this verification procedure; however, formalizing this concept requires more work (see \cref{sec:q-Money}).

What is more, the technological requirements of conjugate coding are very basic: the single-qubit ``prepare-and-measure'' paradigm of conjugate coding is feasible with today's technology---thus, many protocols derived from conjugate coding inherit this desirable property (which is, in fact considered the \emph{gold standard} for ``feasible'' quantum protocols).


\label{sec:q-Money}

In the late 1960's, Wiesner~\cite{Wie83} had the visionary idea that quantum information could be used to create unforgeable bank notes. His ideas were in fact so much ahead of their time that it took years to publish them! (According to~\cite{BBB14}, Wiesner's original manuscript was written in 1968.)

In a nutshell, Wiesner's proposal consists in  \emph{quantum banknotes}  created by encoding quantum particles using conjugate coding (\cref{sec:conj-coding}), with both the classical information and basis choice being chosen as random bitstrings. Thus, a banknote consists of a sequence of single qubits, chosen randomly from the states $\{\ket{\updownarrow}, \ket{\leftrightarrow}, \ket{\neswarrow}, \ket{\nwsearrow}\}$.
As discussed in \cref{sec:conj-coding}, the originator of the quantum banknote (typically called ``the bank'') can verify that a quantum banknote is genuine, yet quantum mechanics prevents essentially any possibility of counterfeiting.
Clearly such a functionality is beyond what classical physics can offer: since any digital record can be copied, classical information simply cannot be used for uncloneability (not even computational assumptions will help).

Wiesner's work was improved and extended in many ways: early work of Bennett, Brassard, Breidbart and Wiesner~\cite{BBBW82} showed how to combine computational assumptions with conjugate coding in order to achieve a type of \emph{public verifiability} for the encoded states (they coined their invention \emph{unforgeable subway tokens}). Further work on publicly-verifiable (also called \emph{public-key} quantum money) includes schemes based on the computational difficulty of some knot-theory related problems~\cite{FGH+12} (see also~\cite{AFG+12}), verification ``oracles''~\cite{Aar09} and hidden subspaces~\cite{AC12}.

Returning to Wiesner's scheme (which is often called \emph{private-key} quantum money in order to distinguish it from the public-key quantum money schemes), we note that the first proof of security in the case of multiple qubits is based on semi-definite programming, and appeared only recently~\cite{MVW13} (this result is tight, since it also gives an explicit \emph{optimal} attack).
We also note work on variants of Wiesner's scheme in which quantum encodings are returned after validation: in all cases (whether the post-verification state is always returned~\cite{FGH+10},
or the post-verification state is returned \emph{only for encodings that are deemed valid}~\cite{BNSU14arxiv}), the resulting protocol has been found to be insecure.

We also note that further work has studied the possibility of private-key quantum money that can be verified using only \emph{classical} interaction with the bank~\cite{Gav12,MVW13},  quantum \emph{coins}~\cite{MS10} (which provide a perfect level of anonymity), as well as noise-tolerant versions of Wiesner's scheme~\cite{PYJ+12}.


\subsection{Quantum Key Distribution}
\label{sec:QKD}
Quantum key distribution (QKD) is by far the most successful application of quantum information to cryptography. By now, QKD is the main topic of a large number of surveys
(see, for instance, \cite{Ben92,BC96,GRTZ02,BEM+07,Feh10}).
Due to abundance of very good references on this topic, we survey it only briefly here.

The ``BB84'' protocol~\cite{BB84,BB14} was the first to show how conjugate coding could be used for an information-theoretically secure key agreement protocol. In a nutshell, the protocol consists in Alice sending a sequence of single qubits, chosen randomly from the states $\{\ket{\updownarrow}, \ket{\leftrightarrow}, \ket{\neswarrow}, \ket{\nwsearrow}\}$. Bob chooses to measure them according to his own random choice of measurement bases. They communicate their basis choice for each encoded qubit; \emph{eavesdropper detection} is performed by comparing the measurement results on a fraction of the bases on which their choices coincide---if successful, this procedure gives a bound on the secrecy and similarity of the remaining shared string, which can be used to distill an almost-perfect shared secret between Alice and Bob. In order to prevent man-in-the-middle attacks, this procedure requires \emph{authenticated} classical channels. Usually, authentication is achieved by an initial shared classical secret between Alice and Bob. Thus, QKD is more accurately described as a key-expansion primitive. We note that, as a theoretical or experimental tool, it is often useful to consider a protocol equivalent to BB84, where the random choice of encoding basis (rectilinear of diagonal) is \emph{delayed}; thus a quantum source would produce a sequence of maximally entangled states $(\ket{00}+\ket{11})/\sqrt{2}$, with both Alice and Bob then measuring in their random choice of bases. That an entangled system could be used in lieu of single qubits was suggested by Ekert~\cite{Eke91}, but note that Ekert's idea was to base  security  on the observation of a Bell-inequality violation --- which implies a set of different measurements than in the rectilinear/diagonal bases.
The entanglement-based (``purified'') BB84 protocol was introduced by Bennett, Brassard and Mermin in~\cite{BBM92}.

We briefly mention that the formal security of QKD was originally left open, and that a long sequence of works (\emph{e.g.}~\cite{LC99,May01}) culminated in a relatively accessible proof by Shor and Preskill, based on the use of quantum error correction~\cite{SP00}. Further work by Renner~\cite{Ren05} showed a very different approach for proving the security of QKD based on exploiting the symmetries of the protocol (and applying a \emph{de Finetti} style representation theorem), and splitting the security analysis into the information-theoretic steps of error-correction and privacy amplification. Other proofs of QKD are more directly based on the complementarity of the measurements~\cite{Koa09}. It is a sign for the complexity of QKD security proofs that most articles on this topic focus only on subparts of the security analysis and only very recently did a first comprehensive analysis of security  appear~\cite{TL15arxiv}.

The huge success of QKD is due in part to the fact that it is readily realizable in the laboratory (the first demonstration appeared in 1992~\cite{BBB+92}).
In light of practical implementations, security proofs for QKD need to be re-visited in order to obtain concrete security parameters---this is the realm of \emph{finite-key security}~\cite{SR08,TLGR12,HT12,TL15arxiv}. Furthermore, we note that when it comes to real-world implementations, QKD is vulnerable to \emph{side-channel} attacks, which are due to the fact that physical implementations deviate from the idealized models used for security proofs (this is often referred to as \emph{quantum hacking}~\cite{LWW+10}).

We further note that the assumption of an initial short shared secret (for authenticating the classical channel) in the implementation of QKD can be replaced with a computational assumption or an assumption about the storage capabilities of the eavesdropper (see \cref{sec:Bounded-storage}). The result is \emph{everlasting}~\cite{Unr13} or \emph{long-term}~\cite{SML10} security: information-theoretic security is guaranteed
\emph{except} during the short period of time during which we assume a computational (or memory) assumption holds.


\subsection{Bit Commitment  implies Oblivious Transfer}
\label{sec:BC=>OT}

Oblivious transfer (OT) and bit commitment (BC) are two basic and important primitives for cryptography. In the classical case, it is easy to show that OT implies BC (in the information-theoretic setting), but the implication in the other direction does not hold.\footnote{See~\cite{Unr13}[Lemma~7] for a proof in the UC framework; the stand-alone impossibility is folklore and can be derived from the impossibility of OT in the plain model~\cite{Lo97} (see also \cref{sec:2PartyImpossible}).}
In stark contrast, OT and BC are known to be \emph{equivalent} in the quantum world. In the following sections, we introduce these primitives (\cref{sec:OTBCdef}) and describe a quantum reduction from oblivious transfer to bit commitment (\cref{sec:quantum-OT}).

\subsubsection{Oblivious Transfer (OT) and Bit Commitment (BC)} \label{sec:OTBCdef}
Wiesner's  paper about quantum cryptography~\cite{Wie83}  introduced ``a means for transmitting two messages either but not both of which may be received''. This classical cryptographic primitive was later rediscovered (under a slightly different form) by  Rabin~\cite{Rab81}, and was given in the form of  \emph{1-out-of-2 Oblivious Transfer (OT)}) by  Even, Goldreich and Lempel~\cite{EGL85}. In
OT, Alice sends two messages $m_0,m_1$ to Bob who receives only one of the messages $m_c$ according to his choice bit $c$. Security for Alice (against dishonest Bob) guarantees that Bob receives only one of the two messages, whereas security for Bob (against dishonest Alice) ensures that Alice does not learn anything about Bob's choice bit\footnote{In fact, formalizing these innocent-looking requirements correctly turns out to be rather tricky~\cite{CSSW06,FS09}}. In the version by Rabin~\cite{Rab81}, this primitive is essentially a secure erasure channel where Alice sends a single bit to Bob. This bit gets erased with probability~1/2 (in this case Bob receives~$\bot$), but Alice does not learn whether the bit was erased. In fact, it is known that Rabin OT is \emph{equivalent} to 1-out-of-2  OT~\cite{Cre88}.

The importance of OT is embodied by the fact that it is \emph{universal} for secure two-party computation~\cite{Kil88} (\emph{i.e.}~using several instances of 1-out-of-2 OT, any function can be securely evaluated among two parties such that no dishonest player can learn any information about the other player's input---beyond what can already be inferred from the output of the computed function).\footnote{A prime example of a secure two-party computation is Yao's millionaire's problem~\cite{Yao82}: two millionaires want to compare their fortune without telling the other specifically how much money they own. This problem can be solved by the secure computation of the greater-than function.} Due to this universality, the innocent-looking OT primitive gives an excellent indicator for the cryptographic power of a model.

\emph{Bit Commitment (BC)} is a cryptographic primitive that captures the following two-party functionality:
Alice has a bit~$b$ that she wants to commit to Bob, but she wants to prevent Bob from reading~$b$ until she chooses to reveal it (\emph{concealing} or \emph{hiding}).
Although Bob should not be able to determine~$b$ before Alice reveals it, Alice should be unable to change the bit after it is committed (\emph{binding}).
A physical-world implementation of bit commitment would be for Alice to write~$b$ on a piece of paper, lock it in a safe, and send the safe to Bob. Since Bob cannot open the safe, he cannot determine~$b$ (concealing), and since Alice has physically given the safe to Bob, she cannot change~$b$ after the commitment phase (binding).  When Alice wishes to reveal the bit, she sends the key to Bob.

\subsubsection{Quantum Protocol for Oblivious Transfer}
\label{sec:quantum-OT}
In~\cite{BBCS92}, Bennett, Brassard, Cr{\'e}peau, and Skubiszewska suggested a very natural quantum protocol for OT (assuming BC): suppose Alice would like to obliviously send $m_0$ and $m_1$ so that Bob receives the message $m_c$ according to his choice bit $c$. She uses conjugate coding to send $n$
quantum states each chosen randomly from the states $\{\ket{\updownarrow}, \ket{\leftrightarrow}, \ket{\neswarrow}, \ket{\nwsearrow}\}$ to Bob. Let us denote by $x \in \{0,1\}^n$ the string of encoded bits and by $\theta \in \{R,D\}^n$ the string of basis choices. Bob measures the received qubits in a random basis $\theta' \in \{R,D\}^n$ of his choice, resulting in outcomes $x' \in \{0,1\}^n$. After Alice tells Bob the bases~$\theta \in \{R,D\}^n$ she was using, Bob can partition the set of indices into two disjoint sets $I_0 \cupdot I_1 = \{1,2,\ldots,n\}$: depending on his OT choice bit $c$, he puts all the indices where he measured correctly in $I_c$ and the rest in $I_{1-c}$. Bob then informs Alice about $I_0,I_1$ (in this fixed order, independent of $c$). Alice picks two independent hash functions $f_0,f_1$ (mapping from $n/2$ bits to 1~bit) and sends $s_i = f_i(x|_{I_i}) \oplus m_i$ for $i=0,1$ to Bob. Here, $x|_I$ denotes the substring of~$x$ with bit indices in $I$. Bob will be able to recover $m_c$ by computing $f_c(x'|_{I_c}) \oplus s_c$.

While it is easy to show that the above protocol is correct and secure against dishonest Alice (\emph{i.e.}~Alice does not learn anything about Bob's choice bit $c$), it is clearly insecure against a dishonest Bob who is able to store all quantum states until Alice tells Bob the basis string $\theta$. Such a Bob can then measure all positions in the correct basis and hence recover both $m_0$ and $m_1$. The idea of~\cite{BBCS92} was to \emph{force} Bob to perform the measurement by requiring him to commit to the bases~$\theta'$ and outcomes~$x'$. Alice then checks a fraction of these commitments \emph{before} Alice announces the basis string $\theta$.

A long line of research \cite{CK88,MS94,Yao95,May96b,BBCS92} has worked towards proving the security of this protocol. However, the crucial tools for an actual proof were eventually developed by Damg\r{a}rd, Fehr, Lunemann, Salvail, and Schaffner~\cite{DFL+09} nearly two decades after the original protocol was proposed; 
Unruh subsequently used these techniques to formally establish the equivalence of BC and OT in the quantum UC model~\cite{Unr10}.


\subsection{Limited-Quantum-Storage Models}
\label{sec:Bounded-storage}
As we will see in \cref{sec:impos-bit-commitment}, bit commitment is impossible to construct in the quantum world. More generally, it has been shown (see \cref{sec:2PartyImpossible}) that secure two-party computation is impossible in the plain quantum model, without any additional restrictions on the adversaries. One option in order to obtain security is to make computational assumptions. However, as we discuss below, it is also possible to obtain information-theoretic security, while making instead some reasonable assumptions about the storage capabilities of the adversary.

One of the challenges in building quantum devices is the difficulty of storing quantum information in a physical system
(such as atomic or phototonic systems) under stable conditions over a long period of time---building a reliable quantum memory is
a major research goal in experimental quantum physics (see e.g. \cite{SAA+10} for a review produced by
the European integrated project \emph{Qubit Applications (QAP)}). Despite continuous progress over the last years,
large-scale quantum memories that can reliably store quantum information 
are
currently out of reach. As we discuss in this section, ingenious quantum cryptographers have turned this technological challenge into an advantage for quantum cryptography!

The \emph{bounded-quantum-storage} model, introduced by Damg\r{a}rd, Fehr, Salvail and Schaffner in~\cite{DFSS05}, is a model which assumes that an adversary can only store a limited number of qubits. Generally, protocols in this model require no quantum storage for the honest players, and are secure against adversaries that are unable to store a constant fraction of the qubits sent in the protocol.\looseness=-1

This model is inspired by the classical bounded-storage model, as introduced by   Maurer~\cite{Mau92,CM97}. In this model, honest parties are required to store $\Theta(n)$ bits, but protocols
(for OT and key agreement) are insecure against attackers with storage capabilities of $\Omega(n^2)$ bits.
Unfortunately, this gap between storage requirements for honest and dishonest players can never be bigger than quadratic~\cite{DM04,DM08}. Combined with the fact that classical storage is constantly getting smaller and cheaper, this quadratic gap puts the classical-bounded-storage assumption on a rather weak footing. In sharp contrast,
the \emph{quantum} bounded-storage model gives an unbounded gap between the quantum-storage requirements of the honest and dishonest players, making this model
model  robust to technological improvements.

In the bounded-quantum storage model, a protocol for OT was proposed~\cite{DFSS05}. Again, it is based on conjugate coding and is essentially identical to the protocol outlined in the previons \cref{sec:quantum-OT}, except that there is a waiting time $\Delta t$ (say, 1 second) right after the quantum phase, before Alice sends her basis string $\theta$ to Bob. In this time, a dishonest receiver Bob is forced to use his (imperfect) quantum memory and therefore loses some information about Alice's string $x$ which intuitively leads to the security of the oblivious transfer. In a subsequent series of works~\cite{DFR+07,DFSS07,DFSS08}, protocols for BC, OT and password-based identification (\emph{i.e.}\ the secure evaluation of the equality function) were presented. For an overview of these results, see~\cite{Sch07,Feh10}.

The \emph{noisy-quantum-storage} model, as introduced by Wehner, Schaffner and Terhal~\cite{WST08} captures  the difficulty of storing quantum information more realistically.
Whereas in the bounded-quantum-storage model, the \emph{physical number of qubits} an attacker can store is limited, dishonest players are allowed arbitrary (but imperfect) quantum storage in the noisy-quantum-storage model.

\paragraph{Beyond Limited Quantum Storage.}
Continuing the idea of assuming realistic technological restrictions on the adversary, researchers have developed protocols that are secure under the assumption that certain classes of quantum operations are hard to perform. A natural class to study consists of adversaries who can store perfectly all qubits they receive, but who cannot perform any quantum operations, except for single-qubit measurements (adaptively in arbitrary bases) at the end of the protocol. Such a model was first studied by Salvail in~\cite{Sal98}, and later by Bouman, Fehr, Gonz\'ales-Guill\'en and Schaffner~\cite{BFGS13} and Liu~\cite{Liu14a,Liu14b,Liu15} under the name of ``isolated-qubit model''.

\paragraph{Cryptographic Proof Techniques.}
In \cref{sec:conj-coding}, we mentioned \emph{uncertainty relations} (and in particular \emph{entropic} uncertainty relations---which quantify uncertainty in information-theoretic terms). These relations play a key role in the security proofs for protocols in the limited-quantum-storage model. We refer to~\cite{WW10} for a survey by Wehner and Winter on this topic. In fact, one can argue that the areas of limited-quantum storage models and entropic uncertainty relations have benefited a lot from each other, as research questions in one area have led to results in the other and vice versa. This fruitful co-existence is witnessed by a series of publications:~\cite{NBW12,BFGS13,BBCW13,BFW14,DFW15}.

\paragraph{Composability}
It is natural to ask whether limited-quantum-storage protocols for basic tasks such as OT can be composed to yield more involved two- or multi-party secure computations. This question was answered in the positive in a number of works, including: Fehr and Schaffner~\cite{FS09}, Wehner and Wullschleger~\cite{WW08} (for sequential composition) and  Unruh~\cite{Unr11} (for bounded concurrent composition).

\paragraph{Implementations.} The technological requirements to implement limited-quantum-storage protocols in practice are modest and rather similar to already available QKD technology (often, the actual quantum phase is the same as in QKD). A small but significant difference is that it makes sense to run secure computations among players which are located within a few meters of each other, whereas the task of distributing keys demands large separations between players. This difference allows experimenters to optimize some parameters (such as the rate) differently for secure-computation protocols. The experimental feasibility of these protocols was analyzed theoretically in~\cite{WCSL10} and demonstrated practically in~\cite{NJM+12,ENG+14}.


\subsection{Delegated Quantum Computation}
\label{sec:delegatedQC}
Quantum computers are known to enable extraordinary computational feats unachievable by today's devices~\cite{Sho94,Gro96,Mos09}.
However, technologies to build quantum computers are currently in their infancy; the current state-of-the-art suggests that, when quantum computers become a reality, these devices are likely to be available at a few location only.
In this context, we envisage the outsourcing of quantum computations from quantum computationally weak clients to universal quantum computers (a type of \emph{quantum cloud} architecture). This scenario has appealing cryptographic applications, such as the delegated execution of Shor's algorithm~\cite{Sho94} for factoring, and thus breaking RSA public keys~\cite{RSA78}.
 From the cryptographic point of view, this scenario raises many questions in terms of the possibility of privacy in delegated quantum computation.

Pioneering work of Childs~\cite{Chi05} and Arrighi and Salvail~\cite{AS06} studied this problem for the first time.
The first practical and universal protocol for private delegated quantum computation, called ``universal blind quantum computation'' (uBQC) was given by Broadbent, Fitzsimons and Kashefi~\cite{BFK09}. In uBQC, the client only needs to be able to prepare random single-qubit auxiliary states (the client requires no quantum memory or quantum processor). Via a classical interaction phase, the client remotely drives a quantum computation of her choice, such that the quantum server cannot learn any information about the computation that is  performed---with only the client learning the output. The uBQC protocol has been demonstrated experimentally~\cite{BKB+12}.

It is remarkable that uBQC is \emph{also} based on conjugate coding! For the first time, it is an application where the states derived from conjugate coding are used to directly achieve \emph{computational} cryptographic tasks (versus other applications of conjugate coding which essentially directly measure these states in order to extract classical information). This relationship with conjugate coding is more clearly apparent in a related protocol called ``quantum computing on encrypted data'' (QCED)~\cite{FBS+14,Bro15}. Here, the computation (as given by a quantum circuit) is public, but is executed remotely on an \emph{encrypted} version of the data (reminiscent of the work on classical \emph{fully homomorphic encryption} \cite{Gen09,RAD78}). In this situation, QCED shows that it is possible to achieve delegated quantum computation where the client only needs to send random states in $\{\ket{\leftrightarrow},\ket{\updownarrow}, \ket{\nwsearrow},\ket{\neswarrow}\}$ (hiding of the computation itself can be achieved via a universal circuit construction).

We mention further that the \emph{verifiability} of delegated quantum computations has been addressed in~\cite{ABE10,FK12arxiv,BGS13,KDK15}, and that the protocol of \cite{FK12arxiv} has been the object of an experiment~\cite{BFKW13}. Also, security of delegated quantum computation has been analyzed in terms of a strong notion of~\emph{composability}~\cite{DFPR14}. Furthermore, work of Reichardt, Unger and Vazirani shows that delegated quantum computation is achievable for a purely classical client, if we are willing to make the assumption of \emph{two} universal quantum computers that cannot communicate~\cite{RUV13} (see also \cref{sec:deviceindependence}, as well as recent work~\cite{GKW15,HPF15arxiv} that improves on this result using uBQC).


\subsection{Quantum Protocols for Coin Flipping and Cheat-Sensitive Primitives}
\label{sec:weak-coin-flip}

In a classic cryptography paper~\cite{Blu83}, Blum describes how to ``flip a coin over the telephone'' with the help of bit commitment: Alice commits to a random bit $a$, Bob tells Alice another random bit~$b$, and Alice opens the commitment to~$a$. The outcome of the coin is $a \oplus b$ which cannot be biased by any of the two players (intuitively, because at least one random bit of an honest player was involved in determining the outcome). A coin flip with this property is called a \emph{strong} coin flip. In contrast, for a weak coin flip, Alice and Bob have a desired outcome, \emph{i.e.}\ Alice ``wins'' if the outcome is 0, and Bob ``wins'' if the outcome is~1. A weak-coin-flipping protocol with bias~$\eps$ guarantees that no dishonest player can bias the coin towards his or her desired outcome with probability greater than~$\eps$. In the classical world, coin-flipping can be achieved under computational assumptions. However, in the information-theoretic setting, it was shown~\cite{HMU06,HW11}  that one of the players can always achieve his desired outcome with probability 1.

In the quantum world,  we  note  that the  general impossibility results for quantum two-party computation (\cref{sec:2PartyImpossible}) are not applicable to coin flipping, since the participants in a coin flipping protocol have no inputs, and instead aim to implement a \emph{randomized} functionality. Nevertheless,
Kitaev showed~\cite{Kit03} (see also~\cite{ABDR04}) that any quantum protocol for strong coin-flipping is insecure since it can be biased by a dishonest player. Formally, the bias of any strong coin-flipping protocol is bounded from below by $\frac{1}{\sqrt{2}}-\frac12$.
Interestingly, Mochon~\cite{Moc07} managed to expand Kitaev's formalism of \emph{point games} to prove the existence of a \emph{weak} coin-flipping protocol with arbitrarily small bias $\eps >0$.
Unfortunately, Mochon's 80-page proof has never been peer-reviewed and is rather difficult to follow. Aharonov, Chailloux, Ganz, Kerenidis and Magnin~\cite{ACG+14arxiv} have managed to simplify this proof considerably.

Based on this result, Chailloux and Kerenidis~\cite{CK09}  derived an optimal strong-coin-flipping protocol with the best possible bias $\frac{1}{\sqrt{2}} - \frac12$, matching Kitaev's lower bound.
Also based on a weak-coin flip, Chailloux and Kerenidis~\cite{CK11} gave the best possible imperfect quantum bit commitment. For the optimality, they prove that in any quantum bit commitment protocol, one of the players can cheat with significant probability\footnote{Formally, $\max\{P_A^*,P_B^*\} \geq 0.739$ where $P_A^*$ is the average over the probabilities that a dishonest committer Alice successfully reveals bit $b=0$ and successfully reveals $b=1$; and $P_B^*$ is the probability that a dishonest verifier Bob guesses the committed bit $b$ after the commitment phase. $P_A^*=P_B^*=\frac12$ holds for a perfect bit-commitment protocol.}. Such a result shows that an imperfect bit commitment cannot be amplified to a perfect one---which severely limits the applicability of the scheme to the cryptographic setting.

\paragraph{Cheat Sensitivity.}
Quantum mechanics offers the possibility to construct imperfect cryptographic primitives in the sense that they are correct (as long as the players are honest), but they are insecure, \emph{i.e.\ }they do allow one of the players (say Alice) to cheat. However, the other player Bob has the possibility to check if Alice has been cheating (possibly by sacrificing the protocol outcome he would have obtained if he followed the protocol without checking). Hence, a cheating Alice has non-zero probability to be detected. These protocols are called \emph{cheat sensitive}~\cite{ATVY00,HK04,BCH+08,GLM08,GLM10,JSG+11,CLM+14}.
In this context, it is argued that one could set up a game-theoretic environment: a player caught cheating has to pay a huge fine (or undergo another punishment) and is therefore deterred from actually doing it.

We note, however, that the applications of cheat sensitive protocols to the cryptographic setting are limited: while quantum protocols for imperfect and cheat-sensitive primitives can provide nice examples of separations between the classical and quantum worlds, they fulfill their purpose as long as they are considered as ``final products'', for instance in case of private information retrieval. However, it is difficult to argue that a strong coin flip with a constant bias, an imperfect bit commitment, or imperfect OT are cryptographically useful primitives, because they do not inherit the cryptographic importance of their perfect counterparts which can be used as building blocks for more advanced cryptographic primitives. In the case of cheat sensitivity, it is often unclear how such primitives behave under composition. In fact, it is a challenging open question to come up with a composability framework for cheat-sensitive quantum primitives.


\subsection{Device-Independent Cryptography} \label{sec:deviceindependence}

An exciting feature of quantum cryptography is that it allows the possibility of \emph{device-independent cryptography} in the sense that protocols can be run on untrusted devices which have possibly been constructed by the adversary. The crucial insight is that the ``quantumness'' of two (or more) devices can be tested and guaranteed by using the devices to violate a Bell inequality, \emph{i.e.}\ to produce correlations that are stronger than allowed by classical mechanics. As outlined in \cref{sec:basics-entanglement}, the most well-known example of such an inequality is the CHSH game~\cite{CHSH69}.
The key observation of device-independent cryptography is that in order to violate the CHSH inequality, a certain amount of intrinsic quantum randomness has to be present in the players' outputs. That we could exploit this relationship for cryptography was originally pointed out by Ekert~\cite{Eke91}, and further studied by
Mayers and Yao~\cite{MY98} and Barrett, Hardy and Kent~\cite{BHK05}. In fact, this latter work shows not only how to accomplish cryptography with untrusted devices,  but also how to do away completely with assumptions on the validity of quantum mechanics: instead, it shows how to accomplish QKD solely based on the non-signaling principle~\cite{HRW10,MPA11}!

The relation between the CHSH violation and the amount of entropy in the outcomes of the measurements can be quantified exactly~\cite{PAM+10}. In fact, on the topic of 
\emph{self-testing quantum devices}~\cite{MY04,MMMO06,MYS12,MS13}, Reichardt, Unger and Vazirani have shown a strong robustness result~\cite{RUV13} in the sense that being close to winning the CHSH-game with optimal probability implies that the players must essentially be in possession of a state which is close to an EPR pair. This is an extremely powerful result which has various applications.

The two qubits of an EPR state are maximally entangled. Quantum mechanics forbids any third party to be entangled with such a state (a phenomenon called \emph{monogamy of entanglement}). Hence, measurements on an EPR state result in shared randomness which is guaranteed to be unknown to any eavesdropper.\footnote{In fact, the first formal security proof of QKD by Shor and Preskill~\cite{SP00} exploited this monogamy of entanglement by showing that a QKD protocol can be transformed (in a series of steps) into a protocol that distills pure EPR states which Eve cannot be entangled with. These EPR states are then measured by Alice and Bob to obtain the secure shared key.} In a similar vein, one can argue that the measurement outcomes of Alice and Bob while successfully playing the CHSH game cannot be known to any adversary \emph{even if this adversary has built the devices herself and is possibly still entangled with them}.

This effect leads to the interesting cryptographic applications of \emph{device-independent randomness amplification and expansion} and \emph{device-independent quantum key distribution}. In \emph{randomness amplification}, the task at hand is to obtain near-perfect randomness from a weak random source using untrusted quantum devices (without using any additional randomness); this idea was originally proposed by Colbeck and Renner~\cite{CR12}.
In \emph{randomness expansion}, one wants to expand a few truly random bits into more random bits, again using untrusted quantum devices; this idea was originally proposed by Colbeck and Kent~\cite{Col06,CK11}. Providing formal security proofs has turned out to be rather challenging and was first established against classical adversaries~\cite{PAM+10,PM13,FGS13}, and later also against quantum adversaries~\cite{VV12,MS14}. A combination of the latest protocols allows to arbitrarily amplify very weak sources of randomness in a device-independent fashion. 
Experimental realizations of device-independent randomness include~\cite{PAM+10,CMA+13}.

In device-independent quantum key distribution, we make the additional assumption that there is no communication between the adversary and the quantum devices.
The first formal proof for a device-independent quantum key distribution scheme was given by Vazirani and Vidick~\cite{VV14}. Current research in this area aims to propose more practical device-independent QKD schemes that retain their functionality at realistic levels of noise.



\section{Quantum Cryptographic Limitations and Challenges}
\label{sec:limitations}

In this section, we survey a number of limitations and challenges of quantum cryptography
(see \cref{sec:intro-summary} for a brief overview of these topics). We cover the impossibility of information-theoretically secure quantum bit commitment (\cref{sec:impos-bit-commitment}) as well as the impossibility of information-theoretically secure two-party quantum computation~(\cref{sec:2PartyImpossible}). Next, we survey the challenges imposed by quantum information in the context of \emph{quantum rewinding} (\cref{sec:quantum-rewinding}) and superposition access to oracles in a quantum world~(\cref{sec:quantum-random-oracles}). Finally, we discuss impossibility results for  position-based quantum cryptography (\cref{sec:position-based}).

\subsection{Impossibility of Quantum Bit Commitment}
\label{sec:impos-bit-commitment}
The ten-year period following the publication of the first quantum key distribution protocol~\cite{BB84} saw only a handful of cryptographers working in quantum cryptography. This era was a period of vivid optimism.
Indeed, the concept that quantum mechanics could allow unconditionally secure key expansion is mind-boggling, so  why stop there? The next natural step to examine was \emph{oblivious transfer}, which is an important building block for cryptography~\cite{Kil88} (see \cref{sec:OTBCdef} for definitions of \emph{bit commitment} and \emph{oblivious transfer}).

From this early period of quantum cryptography, we know of a quantum reduction from bit commitment to oblivious transfer~\cite{BBCS92}
(see \cref{sec:BC=>OT}).
Hence, the holy grail of oblivious transfer is achievable, \emph{if only we have access to a bit commitment}! Thus, researchers explored the possibility of  \emph{quantum bit commitment} (\emph{i.e.}\ of using quantum information in order to build bit commitment), with the hopes of founding all of cryptography on the unique assumption of quantum mechanics. This line of work started in~\cite{BC91}, culminating in a claim of a unconditionally secure quantum bit commitment protocol~\cite{BCJL93}.
However, the optimism for quantum cryptography lasted only a few years as Mayers~\cite{May96}
found a subtle flaw in the original argument of security. This result was generalized to rule out all quantum protocols for bit commitment by Mayers, and Lo and Chau~\cite{May97,LC97}. Note that the \emph{possibility} of bit commitment in the \emph{limited-quantum-storage} model (\cref{sec:Bounded-storage}) introduces an extra physical assumption, and does not contradict the impossibility as discussed here!

We now briefly review the main impossibility argument~\cite{BCMS97arxiv} (for ease of presentation, we focus on the exact case)\footnote{See also a more recent proof exploiting ``quantum combs''~\cite{CDP+13}, as well as a more general result about the impossibility of ``growing'' quantum bit commitments~\cite{WTHR11}.}. First, consider the following sketch of impossibility for perfectly secure \emph{classical} bit commitment: suppose such a protocol exists. Then by the information-theoretic security requirement, at the end of the commitment phase, Bob's view of the protocol must be independent of~$b$ (since, otherwise, the protocol would not be perfectly hiding). But this independence implies that Alice can choose to reveal either $b=0$ or~$b=1$ in the reveal phase, with both being accepted by Bob. Hence, the bit commitment cannot be binding. It is interesting
that the same proof structure is applicable to the quantum case, albeit by invoking some slightly more technical tools. Namely, we first consider a \emph{purified} version of the protocol, which consists in all parties acting at the quantum level (measurements are replaced by a unitary process via a standard technique). Next, by the information-theoretic hiding property, the reduced quantum state that Bob holds at the end of the commit phase must be identical, whether~$b=0$ or $b=1$. This condition is enough to break the binding property, since it means~\cite{Uhl76} that Alice can locally perform a unitary quantum operation on her system in order to re-create a joint state consistent with either~$b=0$, or $b=1$, at her choosing\footnote{Technically, Uhlmann's theorem states that any two purifications of Bob's reduced quantum state are related by a unitary transform.}. Hence, she can chose to open either~$b=0$ or $b=1$ at a later time, and Bob will accept: the commitment scheme cannot be binding.

Going back to the original paper on quantum bit commitment~\cite{BCJL93}, we note that a subtlety in the definition of the \emph{binding} property is the origin of the false claim of security: while it is true that the protocol is such that Alice is unable to \emph{simultaneously} hold messages that would unveil a commitment to $b = 0$ \emph{and} as $b = 1$ (and thus, to be able to choose to open $b = 0$ \emph{and} $b = 1$), this is insufficient to prove security, since in fact Alice is able to \emph{delay} her choice of commitment until the very end of the protocol---at which point she can \emph{choose} to  open as either $b=0$ or $b=1$ (but not necessarily both at the same time!).

\subsection{Impossibility of Secure Two-Party Computation using Quantum Communication}
\label{sec:2PartyImpossible}
Given the impossibility of quantum bit commitment, the next question to ask is: are there any classical primitives that may be implemented securely using quantum communication? In fact, the possibility for OT was stated as a open problem in~\cite{BC96}. Unfortunately, this hope was shattered rather quickly, as  impossibility results were given by Lo in~\cite{Lo97} for one-sided computations (where only one party receives output). This result already shows the impossibility of 1-out-of-2 OT---the proof technique follows closely the technique developed for the impossibility of quantum bit commitment (see \cref{sec:impos-bit-commitment}).

It took almost ten years until Colbeck showed the first impossibility result for two-sided computations, namely that Alice can always obtain more information about Bob's input than what is implied by the value of the function~\cite{Col07}. In a similar vein, Salvail, Schaffner and Sotakova proved in~\cite{SSS14} that any quantum protocol for a non-trivial primitive leaks information to a dishonest player. What is worse, even with the help of a trusted party, the cryptographic power of any primitive cannot be ``amplified'' by a quantum-communication protocol.

Buhrman, Christandl and Schaffner~\cite{BCS12} have strengthened the above impossibility results by showing that the leakage in any quantum protocol is essentially as bad as one can imagine: even in the case of approximate correctness and security, if a protocol is ``secure'' against Bob, then it is \emph{completely} insecure against Alice (in the sense that she can compute the output of the computation for \emph{all} of her possible inputs). For impossibility results in the universal composability (UC) framework, see~\cite{FKS+13}.

\subsection{Zero-Knowledge Against Quantum Adversaries ---``Quantum Rewinding''}
\label{sec:quantum-rewinding}

Zero-knowledge interactive proofs, as introduced by Goldwasser, Micali, and
Rackoff~\cite{GMR89} are interactive proofs with the property that the verifier learns nothing from her interaction with the honest prover, beyond the
validity of the statement being proved. These proof systems play an important role in the foundations of cryptography, and are also fundamental building blocks to achieve cryptographic functionalities (see~\cite{Gol02} for a survey).

In zero-knowledge interactive proofs, the notion that the verifier ``learns nothing'' is formalized via the \emph{simulation paradigm}: if, for every cheating verifier (interacting in the protocol on a positive instance), there exists a simulator (who does not interact with the prover) such that the output of the verifier is indistinguishable from the output of the simulator, then we say that the zero-knowledge property holds. In the classical world, a common proof technique used for establishing the zero-knowledge property is \emph{rewinding}: a simulator is typically built by executing the given verifier---except that some computation paths are culled if the random choices of the verifier are not consistent with the desired effect. This selection is done by keeping a trace of the interaction, thus, if the interaction is deemed to have followed an incorrect path, the simulation can simply reset the computation (``rewind'') to an earlier part of the computation (see~\cite{Gol02} and references therein).

In the quantum setting, such a rewinding approach is impossible: the no-cloning theorem tells us that it is not possible, in general, to keep a secondary copy of the transcript in order to return to it later on. This problem is further aggravated by the fact that, in the most general case,  the verifier starts with some \emph{auxiliary} quantum information (which we do not, in general, know how to re-create)---thus even a ``patch'' that would emulate the rewinding approach in the simple case would appear to  fail in the case of auxiliary quantum information. We emphasize that the above concerns about the zero-knowledge property are applicable to purely \emph{classical}  protocols: honest parties are completely classical, but we wish to establish the zero-knowledge property against a verifier that may receive, store and process quantum information (these concerns are independent of the \emph{computational power} of the verifier---they simply relate to the computational model!).

The fundamental difficulty in proving the zero-knowledge property in the quantum world was first discussed by van de Graaf~\cite{Gra97}; while some progress was made on this question \cite{DFS04}, it is the breakthrough result of Watrous~\cite{Wat06} that restored confidence that the zero-knowledge property of many standard classical zero-knowledge proofs is maintained in a quantum world.

In a nutshell, Watrous introduced the technique of \emph{quantum rewinding}, which establishes that
under some reasonable (and commonly satisfied) conditions, the
success probabilities of certain processes with quantum inputs and outputs can be
amplified.   This technique therefore provides an alternative to the classical rewinding paradigm, and is used to show that the  Goldreich-Micali-Wigderson graph 3-coloring protocol~\cite{GMW91} is zero-knowledge against quantum attacks.
 We briefly mention that quantum rewinding is established using a technique resembling \emph{amplitude amplification}~\cite{BHMT02} as is related to Grover's quantum search algorithm~\cite{Gro96}.

Further work on quantum rewinding has dealt with extending the domain of applicability to \emph{proofs of knowledge}~\cite{Unr12}.  However,  \cite{ARU14} show limitations to this technique (so that, in fact---relative to an oracle---there exists classical protocols that are insecure against quantum adversaries). See also \cite{DFS04}.

\subsection{Superposition Access to Oracles --- Quantum Security Notions}
\label{sec:quantum-random-oracles}

\emph{Post-quantum cryptography}~\cite{BBD09} (see \cref{sec:further-topics}) investigates classical cryptographic schemes which remain secure in the presence of quantum adversaries. In classical cryptography, security is often defined in terms of an interactive game between an adversary and a challenger: a scheme is deemed secure if the adversary can only win the game with negligible probability. When such notions are used to prove post-quantum security, one must consider quantum adversaries which are potentially able to communicate quantumly with the challenger. An example is the chosen-plaintext-attack (CPA) learning phase that is present in game-based security definitions, for instance for defining indistinguishability (IND) security of encryption schemes~\cite{GM84,KL07}, where it is natural to consider attackers that can query superpositions of plaintexts to be encrypted and are returned superpositions of according ciphertexts from the challenger.

Another important example relates to the random-oracle (RO) model.
 A common technique used in classical cryptography is to assume that hash functions are perfect random oracles which adversaries can evaluate. It is well-known in classical cryptography that the RO-methodology comes with a plethora of   techniques that can be employed in order to give formal proofs. Unfortunately, most of these tricks do not work in a quantum context for the following reason: a quantum adversary can always evaluate a classical hash function on an arbitrary superposition of inputs. Therefore, in the quantum random oracle (QRO) model, it is necessary to give the adversary superposition access to the oracle. As a consequence, standard techniques from the classical RO model (such as planting the challenge in a random one of the RO queries of the adversary) fail in the more realistic QRO model setting (the adversary might make a single quantum query with all input values in superposition).

Boneh, Dagdelen, Fischlin, Lehmann, Schaffner, and Zhandry~\cite{BDF+11} first showed how to correctly define the random-oracle model in the quantum setting. They also showed a separation between the classical and quantum RO models. Zhandry~\cite{Zha12} showed how to plant challenges in the QRO model at the beginning of the execution, and Unruh~\cite{Unr15} showed how to reprogram the RO during runtime. Security definitions allowing superposition access have subsequently been studied by Boneh and Zhandry~\cite{Zha12,BZ13} in the context of encryption, digital signatures and the construction of pseudo-random functions.
See also related work by
Damg{\aa}rd, Funder, Nielsen and Salvail~\cite{DFNS14}, who study superposition attacks on secret-sharing and multi-party protocols.

\subsection{Position-Based Quantum Cryptography}
\label{sec:position-based}

In cryptography, digital keys or biometric features are
used to verify the identity of a person. The goal of position-based
cryptography is to use the \emph{geographical position} of an entity
as a cryptographic credential. As a physical analogy, consider the scenario of a bank, where typically, the mere fact that a bank teller is behind the counter (her \emph{position}) suffices as a credential in order to initiate the exchange of sensitive information.

A central building block of position-based cryptography is the task of
\emph{position verification}, a problem previously studied in the field of wireless security~\cite{BC93,SSW03,VN04,Bus04,CH05,SP05,ZLFW06,CCS06}. The goal is to prove to a set of verifiers that
one is at a certain geographical location.
Protocols typically exploit the relativistic no-signaling principle that messages cannot travel faster than the speed of light. By responding to a verifier in a timely manner, one can guarantee that one is within a certain distance of that verifier~\cite{BC93}. It was shown in~\cite{CGMO09} that classical position-verification protocols based only on this relativistic principle can be broken by multiple attackers who simulate being at the claimed position while physically residing elsewhere in space. Because of the no-cloning property of quantum information (see \cref{sec:no-cloning}), it was believed that with the use of quantum messages one could devise protocols that were resistant to such collaborative attacks. Several schemes were proposed~\cite{KMS11,Mal10a,CFG+10arxiv,Mal10b,LL11} that
later turned out to be insecure. Finally, Buhrman, Chandran, Fehr, Gelles, Goyal, Ostrovsky and Schaffner showed that also in the quantum case, no unconditionally secure schemes are
possible~\cite{BCF+14}, as long as the colluding adversaries share a large enough amount of entanglement: attackers can break the protocol if the number of pre-shared EPR pairs is exponential in the size of the messages of the protocol~\cite{BK11}. This exponential overhead in resources (in terms of entanglement and quantum memory) leads to the main open problem in this research area, namely to find quantum protocols which remain secure under the
assumption that adversarial resources are restricted to a polynomial amount, while at the same
time, honest players can perform the schemes efficiently.

Historically, position-based schemes were first studied by Kent, Monroe and Spiller in 2002 under the name of ``quantum tagging''. A US-patent was granted in 2006~\cite{KMSB06}, but the results appeared  in the scientific literature only in 2010~\cite{KMS11}. Some simple position-verification schemes are studied in~\cite{KMS11,BFSS13}. The only quantum ingredient in these protocols is a single qubit sent to the prover who is required to route this qubit back to the correct verifier depending on the classical information he also receives from the verifiers. Note that the actions of the honest players are simple enough that they can be implemented using current quantum technology.

In order to analyze how much entanglement colluding adversaries need to break these simple schemes, a new model of (classical) communication complexity (called the \emph{garden-hose model}) was introduced by Buhrman, Fehr, Schaffner and Speelman~\cite{BFSS13}. This model connects attacks on position-based quantum protocols to various interesting problems in classical complexity
theory and communication complexity, as witnessed in related work~\cite{CSWX14,KP14}. In particular, the \emph{garden-hose complexity} of a function gives an upper bound on the amount of entanglement required to break the security of the position-verification protocol based on that function. However, it is an open question whether more advanced techniques will allow to also prove lower bounds on the entanglement required to break these simple position-verification protocols.

In~\cite{Unr14b}, Unruh introduces a helpful methodology for analyzing quantum circuits in space-time and gives a position-verification protocol in three dimensions which is secure in the quantum-random-oracle model. Furthermore, \cite{BCF+14,Unr14b}  give schemes for position-based authentication which allows the verifiers to be convinced that a message originated from a certain location. However, it remains an open problem to find efficient schemes which do not use random oracles.



\section{Conclusion and Open Problems}
\label{sec:open-probs}
Since its inception almost 50 years ago, quantum cryptography has developed into an active and exciting multidisciplinary area of research that combines state-of-the-art techniques from cryptography, quantum physics, complexity theory, information theory and beyond. While experimental implementations are still at the prototype level, 
our theoretical understanding of the power and limitations of quantum cryptography is continuously expanding.

Ongoing work on quantum cryptography consists in improving existing schemes as well as finding further applications and proof techniques. As final words, we mention here some open problems of interest.

\begin{itemize}

\item \emph{What types of cryptosystems can quantum algorithms break?} The area of post-quantum cryptography bases classical cryptography on computational problems which are hard even for quantum computers. More research on quantum algorithms for quantum cryptanalysis is needed to fully understand how difficult these problems are. This understanding is also crucial when choosing the security parameters for post-quantum cryptographic schemes.

\item \emph{Can we make device-independent protocols that are feasible in practice?} It is a challenging open problem to develop device-independent protocols (for key distribution, but possibly also for other applications) which can tolerate a realistic amount of noise.

\item \emph{Can quantum protocols verify the position of a player?} As outlined in \cref{sec:position-based}, one of the main open questions in the area of position-based cryptography is to find a protocol which can be executed efficiently (with current technology) by honest players, but requires an exponential amount of resources (such as entangled qubits) for attackers to break it.

\item \emph{How can we construct quantum-secure pseudorandom permutations (qPRP)?} Related to the topic of quantum security notions (\cref{sec:quantum-random-oracles}), Zhandry~\cite{Zha12} has shown how to construct quantum-security pseudo-random functions (qPRF). Classically, it is well-known that using the PRF in a three-round Feistel network yields a pseudo-random permutation. However, this construction is probably insecure in the quantum setting~\cite{KM10}. It is an open question how to construct  quantum-secure pseudo-random permutations.

\item \emph{Which cryptographic functionalities can be achieved by quantum protocols?} The impossibility results from \cref{sec:2PartyImpossible} are all concerned with \emph{deterministic classical} functionalities. In \cref{sec:weak-coin-flip}, we have seen that quantum protocols for (weak or biased strong) coin flipping exist. Hence what exactly is the set of randomized classical functionalities that can be implemented by quantum protocols? More generally, can these impossibility results be extended to quantum functionalities?

\item \emph{Does quantum information allow for devices that hide the inner workings of a computer program?}
Due to its diverse and far reaching applications, program obfuscation has been long considered
as a holy grail of cryptography. However, hopes of attaining highly secure obfuscation were diminished in 2011 by an impossibility proof~\cite{BGI+12} (note, however that weaker security notions are attainable~\cite{GGH+13}). The situation is completely different in the quantum case, since the proof technique is not applicable (essentially due to the no-cloning theorem). As such, a positive result establishing that quantum information allows program obfuscation would unleash a number of powerful primitives, and would yield another qualitative advantage of quantum information over its classical counterpart.

\item \emph{What are the limits of the delegated quantum computation scenario?} In \cref{sec:delegatedQC}, we reviewed results on how a quantum computationally weak client can outsource a quantum computation. An open question that remains is to establish the ultimate limits in terms of the power of the client: can a fully classical client delegate a private quantum computation to a single quantum server, while ensuring privacy and/or verifiability? In the computational setting, this question is related to that of \emph{quantum fully homomorphic encryption}: can we encrypt quantum data such that any quantum circuit can be applied to the encrypted data (without revealing the key, of course!)?

\item \emph{Can we build quantum public-key money from standard assumptions?} Current techniques for quantum public-key money  rely on \emph{ad hoc} assumptions (see \cref{{sec:q-Money}}). An open problem is to construct these primitives on standard cryptographic assumptions (such as the existence of quantum-secure one-way functions~\cite{MRV07arxiv,Ajt96,Pei15eprint}).

\end{itemize}


\section{Acknowledgements}
It is a pleasure to acknowledge conversations with Gilles Brassard, Gus Gutoski,
Serge Fehr and Louis Salvail in the preparation of this
document.
We thank Romain All\'eaume, Matthias Christandl, Claude Cr{\'e}peau, Ivan Damg{\aa}rd, Joseph  Fitzsimons,
Tommaso Gagliardoni, Andreas H{\"u}lsing, Stacey Jeffery, Elham Kashefi, Raza Ali Kazmi,  Anthony Leverrier, Renato Renner, Christoph Simon, Fang Song, Douglas Stebila,
Dominique Unruh, Thomas Vidick, Stefan Wolf, and
Ronald de Wolf  for very useful feedback on an initial version of this document. We also thank Ronald Mullin
for his invitation to submit this paper to the  25th anniversary edition of \emph{Designs, Codes and Cryptography}, and an anonymous reviewer for helpful comments.
We are deeply indebted to Barbara Cai\c{c}ara Santos for help in managing our references database.
A.\,B.'s work is supported by Canada's
NSERC.
C.\,S. is supported by an NWO VIDI grant, and the EU 7th framework project~SIQS.

The bibliographic entries in this survey are available as an open-source \BibTeX\ file at \url{https://github.com/cschaffner/quantum-bib}.

\newcommand{\etalchar}[1]{$^{#1}$}

\end{document}